\documentclass[12pt]{article}
\usepackage{graphicx}
\usepackage{latexsym}
\usepackage{amssymb}
\usepackage{amsmath}

\begin{document}

\makeatletter
\renewcommand*{\@cite}[2]{{#2}}
\renewcommand*{\@biblabel}[1]{#1.\hfill}
\makeatother

\title{Kinematics of B--F Stars as a Function of Their Dereddened Color from Gaia and PCRV Data}
\author{G.~A.~Gontcharov\thanks{georgegontcharov@yahoo.com}}
\date{Pulkovo Astronomical Observatory, Russian Academy of Sciences, Pulkovskoe sh. 65, St. Petersburg, 196140 Russia}

\maketitle

\newpage

Abstract

Parallaxes with an accuracy better than 10\% and proper motions from the Gaia DR1 TGAS
catalogue, radial velocities from the Pulkovo Compilation of Radial Velocities (PCRV), accurate Tycho-2 photometry, 
theoretical PARSEC, MIST, YaPSI, BaSTI isochrones, and the most accurate reddening
and interstellar extinction estimates have been used to analyze the kinematics of 9543 thin-disk B--F stars
as a function of their dereddened color. The stars under consideration are located on the Hertzsprung--Russell diagram 
relative to the isochrones with an accuracy of a few hundredths of a magnitude, i.e., at
the level of uncertainty in the parallax, photometry, reddening, extinction, and the isochrones themselves.
This has allowed us to choose the most plausible reddening and extinction estimates and to conclude
that the reddening and extinction were significantly underestimated in some kinematic studies of other
authors. Owing to the higher accuracy of TGAS parallaxes than that of Hipparcos ones, the median
accuracy of the velocity components $U$, $V$, $W$ in this study has improved to 1.7 km s$^{-1}$, although outside
the range $-0.1^m<(B_T-V_T)_0<0.5^m$ the kinematic characteristics are noticeably biased due to the
incompleteness of the sample. We have confirmed the variations in the mean velocity of stars relative to the
Sun and the stellar velocity dispersion as a function of their dereddened color known from the Hipparcos
data. Given the age estimates for the stars under consideration from the TRILEGAL model and the
Geneva-Copenhagen survey, these variations may be considered as variations as a function of the stellar
age. A comparison of our results with the results of other studies of the stellar kinematics near the Sun has
shown that selection and reddening underestimation explain almost completely the discrepancies between
the results. The dispersions and mean velocities from the results of reliable studies fit into a $\pm2$ 
km s$^{-1}$ corridor, while the ratios $\sigma_V/\sigma_U$ and $\sigma_W/\sigma_U$ fit into $\pm0.05$. 
Based on all reliable studies in the range
$-0.1^m<(B_T-V_T)_0<0.5^m$, i.e., for an age from 0.23 to 2.4 Gyr, we have found: $W=7.15$ km s$^{-1}$,
$\sigma_U=16.0e^{1.29(B_T-V_T)_0}$, $\sigma_V=10.9e^{1.11(B_T-V_T)_0}$, $\sigma_W=6.8e^{1.46(B_T-V_T)_0}$,
the stellar velocity dispersions
in km s$^{-1}$ are proportional to the age in Gyr raised to the power $\beta_U=0.33$, $\beta_V=0.285$, and
$\beta_W=0.37$.

\bigskip\noindent
\leftline {PACS numbers: 97.10.Zr; 98.10.+z; 98.35.Pr}
%

\newpage

\section*{INTRODUCTION}

The dependence of kinematics on astrophysical characteristics of stars has been investigated for a
long time (Parenago 1954, p. 139; Perryman 2009, pp. 302, 490). The dependence of the dispersions
$\sigma_U$, $\sigma_V$, $\sigma_W$ and means $\overline{U}$, $\overline{V}$, $\overline{W}$ 
of the velocity components $U$, $V$, $W$ in the rectangular Galactic coordinate
system
\footnote{$U$ and the $X$ axis are directed toward the Galactic center, $V$ and $Y$ are in the direction 
of Galactic rotation, $W$ and $Z$ are directed toward the North Pole.}
on the intercorrelating color, dereddened color, spectral type, effective temperature, and
age for main-sequence (MS) B--F stars is particularly pronounced.

The quantities $\overline{U}$, $\overline{V}$, $\overline{W}$ for a sample of stars with
specific astrophysical characteristics taken with the opposite sign are considered as the components 
$U_{\odot}$, $V_{\odot}$, $W_{\odot}$ of the solar motion relative to this sample,
i.e., the peculiar solar motion toward the apex (Kulikovskii 1985, p. 74).

The quantities $U$, $V$, $W$ are calculated from the Galactic longitude $l$, latitude $b$, distance $R$, 
or parallax $\varpi$ as well as the proper motion components $\mu_l\cos(b)$ and $\mu_b$
and radial velocity $V_r$ (Kulikovskii 1985, p. 74).
Studies of the dependence of kinematic characteristics on astrophysical ones have reached a new level
of accuracy and completeness after the appearance of the Hipparcos catalogue (ESA 1997) and its new
version (van Leeuwen 2007) with highly accurate positions, parallaxes, and proper motions for tens
of thousands of stars near the Sun, mostly within 200 pc.

Table 1 presents, in chronological order, the key studies of the dependence of these kinematic characteristics
on the dereddened color or characteristics correlating with it for a large number of B--F stars
near the Sun that have appeared after Hipparcos. These do no include the studies where kinematic
characteristics are not matched with fairly accurate astrophysical ones. In addition, we do not consider
the studies of separate small groups of stars or stars far from the Sun, because their kinematics can differ
significantly from the overall kinematics of the local group of stars. In all of the studies under consideration
the stellar positions, parallaxes, and proper motions were taken from Hipparcos. In the column
`Sources of $V_r$' the cases of analyzing the kinematics without using any radial velocities are marked as `-'.
A theory of such an analysis was given by Dehnen and Binney (1998), Mignard (2000), and Aumer and
Binney (2009). Three popular catalogues of radial velocities of bright stars were used in the remaining
studies: the Second Catalogue of Radial Velocities with Astrometric Data (CRVAD-2; Kharchenko
et al. 2007), the Pulkovo Compilation of Radial Velocities (PCRV; Gontcharov 2006), and the Geneva-Copenhagen 
survey of the solar neighborhood (GCS; Nordstr\"om et al. 2004; Holmberg et al. 2007, 2009;
Casagrande et al. 2011). The column `Argument' gives the quantity on which the kinematic characteristics
in the corresponding study depends ($(B_T-V_T)$ and $b-y$ were calculated from Tycho-2 (H\o g
et al. 2000) and Stro\"omgren photometry, respectively).
The presence or absence of results for the corresponding characteristic in the study is marked in the
columns `$\sigma_U$', `$\sigma_V$', `$\sigma_W$' and `$U_{\odot}V_{\odot}W_{\odot}$'.

\begin{table*}[!h]
\def\baselinestretch{1}\normalsize\tiny
\caption[]{Key studies of the dependence of kinematic characteristics on astrophysical ones for B--F stars near the Sun
that have appeared after Hipparcos. $\Delta E(B_T-V_T)$ is the correction to the reddening adopted in the corresponding 
study. $X$ denotes $(B_T-V_T)_0$.
}
\label{table1}
\[
\arraycolsep=0.05em
\begin{tabular}{lcccccccc}
\hline
\noalign{\smallskip}
\texttt{Reference} & \texttt{Designation} & \texttt{Sources of} $V_r$ & \texttt{Argument} & $\sigma_U$ & $\sigma_V$ & $\sigma_W$ & $U_{\odot}V_{\odot}W_{\odot}$ & $\Delta E(B_T-V_T)$  \\
\hline
\noalign{\smallskip}
\texttt{G\'omez et al. (\cite[1997]{gomez})} & \texttt{Gray squares} & \texttt{Various} & Age & $+$ & $+$ & $+$ & $-$ & 0 \\
\texttt{Dehnen and Binney (\cite[1998]{db})} & \texttt{Black triangles} & $-$ & $(B_T-V_T)$ & $+$ & $-$ & $-$ & $+$ & $0.04+0.01X$ \\
\texttt{Mignard (\cite[2000]{mignard})} & \texttt{Gray diamonds} & $-$ & \texttt{Sp. type} & $+$ & $+$ & $+$ & $+$ & 0 \\
\texttt{Aumer and Binney (\cite[2009]{aumer})} & \texttt{Black squares} & $-$ & $(B_T-V_T)_0$ & $+$  & $+$ & $+$ & $-$ & $0.04+0.01X$ \\ 
\texttt{Francis and Anderson (\cite[2009]{francis})} & \texttt{Black diamonds} & \texttt{CRVAD-2} & $(B_T-V_T)$ & $+$ & $-$ & $-$ & $+$ & $0.08-0.05X$ \\ 
\texttt{Gontcharov (\cite[2012c]{apex})} & \texttt{Gray curve} & \texttt{PCRV} & $(B_T-V_T)_0$ & $+$ & $+$ & $+$ & $+$ & $0.24X^3-0.22X^2+0.07X$ \\ 
\texttt{Aghajani and Lindegren (\cite[2013]{aghajani})} & \texttt{Black circles} & \texttt{GCS} & $(b-y)$ & $+$ & $+$ & $+$ & $+$ & $-0.90X^3+1.63X^2-1.00X+0.25$ \\ 
\hline
\end{tabular}
\]
\end{table*}


The symbols representing the results of the studies under consideration in Figs. 1--5, where the kinematic
characteristics are plotted against the dereddened color $(B_T-V_T)_0$ (the recalculation of the original
arguments to $(B_T-V_T)_0$ is discussed below), are given in the column "Designation". The results
of the studies are indicated by separate signs for the subsamples into which the entire sample of stars was
divided in the corresponding study. The results of Francis and Anderson (2009) are shown in Figs. 1a,
2a, 3, and 4 twice: before the exclusion of particular stars from their sample (open black diamonds) and
for the thin-disk stars (filled diamonds). The vertical bar near each sign indicates the formal error of the
corresponding study. For all studies the errors of the dispersions and means of the velocities lie within
the range 0.4--2 km s$^{-1}$ and, in some cases, they are smaller than the size of the sign in the figure.
Figures 2 and 4 differ from Figs. 1 and 3, respectively, in that the corrections to the reddening adopted in the
corresponding study that are discussed below were made in the results of the studies. They are given in
the column `$\Delta E(B_T-V_T)$' of Table 1, where $X$ denotes $(B_T-V_T)_0$. For $\sigma_V/\sigma_U$ and
$\sigma_W/\sigma_U$ the influence of the corrections is small and, therefore, no analog
without corrections is shown for Fig. 5. The thin stepwise curve in Figs. 1, 2, and 5 indicates the kinematic
characteristics adopted in the Besan\c{c}on model of the Galaxy (BMG) (Robin et al. 2003; Czekaj et al. 2014).

The results of this study based on TGAS and PCRV data are also shown in Figs. 1--5. For them,
following G\'omez et al. (1997) and Gontcharov (2012c), we applied a moving averaging. The stars were
arranged by age $(B_T-V_T)_0$ and the sought-for kinematic characteristic was calculated for the subsample
of stars with ordinal numbers from 1 to 500 (the bluest stars), then from 2 to 501, from 3 to 502,
and so on. As a result, a set of similarly averaged values of $(B_T-V_T)_0$ was associated with the set of
values of each kinematic characteristic. The results of Gontcharov (2012c) and this study are indicated
by the gray and black curves, respectively (the gray band $\pm2$ km s$^{-1}$ in width is shown for clarity), while
the results of G\'omez et al. (1997) are indicated by the thin solid black line with gray squares strung on it to
distinguish them from the remaining ones.

The discrepancies between the results of different studies visible in the figures often exceed considerably
the declared formal errors. This study is devoted to explaining these discrepancies. It seems timely,
because the original data for calculating the kinematic characteristics and the arguments on which
they can depend have been updated significantly in recent years. In particular, Gontcharov (2017a) produced
a new 3D reddening map, while Gontcharov and Mosenkov (2017a, 2017b) showed it to be one
of the best maps within a few hundred pc of the Sun. In addition, the theoretical isochrones, for example,
PARSEC (Bressan et al. 2012), MESA Isochrones and Stellar Tracks (MIST) (Paxton et al. 2011; Dotter
2016; Choi et al. 2016), Yale-Potsdam Stellar Isochrones (YaPSI) (Spada et al. 2017), and BaSTI
(Pietrinferni et al. 2004), as well the population synthesis models (stellar population modeling) accompanying
them have been improved compared to their previous analogs. They are already intensively used to
test the dependences of kinematic characteristics on astrophysical ones (Czekaj et al. 2014).

However, the main data updates stem from the appearance of the first results of the Gaia space project
(Gaia 2016) in September 2016. The Gaia DR1 Tycho-Gaia astrometric solution (TGAS) catalogue
(Michalik et al. 2015) with accurate $l$, $b$, $\mu_l\cos(b)$, $\mu_b$ and $\varpi$ for more than two million stars 
from Tycho-2 is among them. In TGAS the parallaxes $\varpi$ were determined
by comparing the measured Gaia positions with the positions of the same stars from Hipparcos and Tycho-2.

In this study we calculated $\sigma_U$, $\sigma_V$, $\sigma_W$, $U_{\odot}$, $V_{\odot}$ and $W_{\odot}$ from the 
TGAS data and PCRV radial velocities and considered their dependences on $(B_T-V_T)_0$. 
The results obtained were compared with the mentioned results of other authors. The following
factors can be responsible for the discrepancies found.
\begin{enumerate}
\item The first and new versions of Hipparcos were used in the studies before and after 2009, respectively,
while TGAS was used in this study.
\item The radial velocities were taken from different sources or were not used at all.
\item Magnitude, distance, etc. constraints were imposed on the sample of stars, which led to various types of selection.
\item A particular astrophysical characteristic was used as an argument. The age calculation from the
observed color by G\'omez et al. (1997) introduces an additional uncertainty into their results.
\item Different reddening estimates were used when calculating the dereddened color from the observed one.
\end{enumerate}

In this study we analyze all these factors and infer the role of each of them.

Note that the parameters of the linear Ogorodnikov-Milne model and the Oort constants $A$, $B$, $C$, and $K$
related to them are not analyzed in this study, because they are calculated for stars near the Sun with a
comparatively low accuracy. In this study, to take into account the Galactic rotation, we adopted 
$A=15.3$ km s$^{-1}$ kpc$^{-1}$ and $B=-11.9$ km s$^{-1}$ kpc$^{-1}$ found by Bovy (2017) for comparatively distant
TGAS stars.

\begin{figure}
\includegraphics{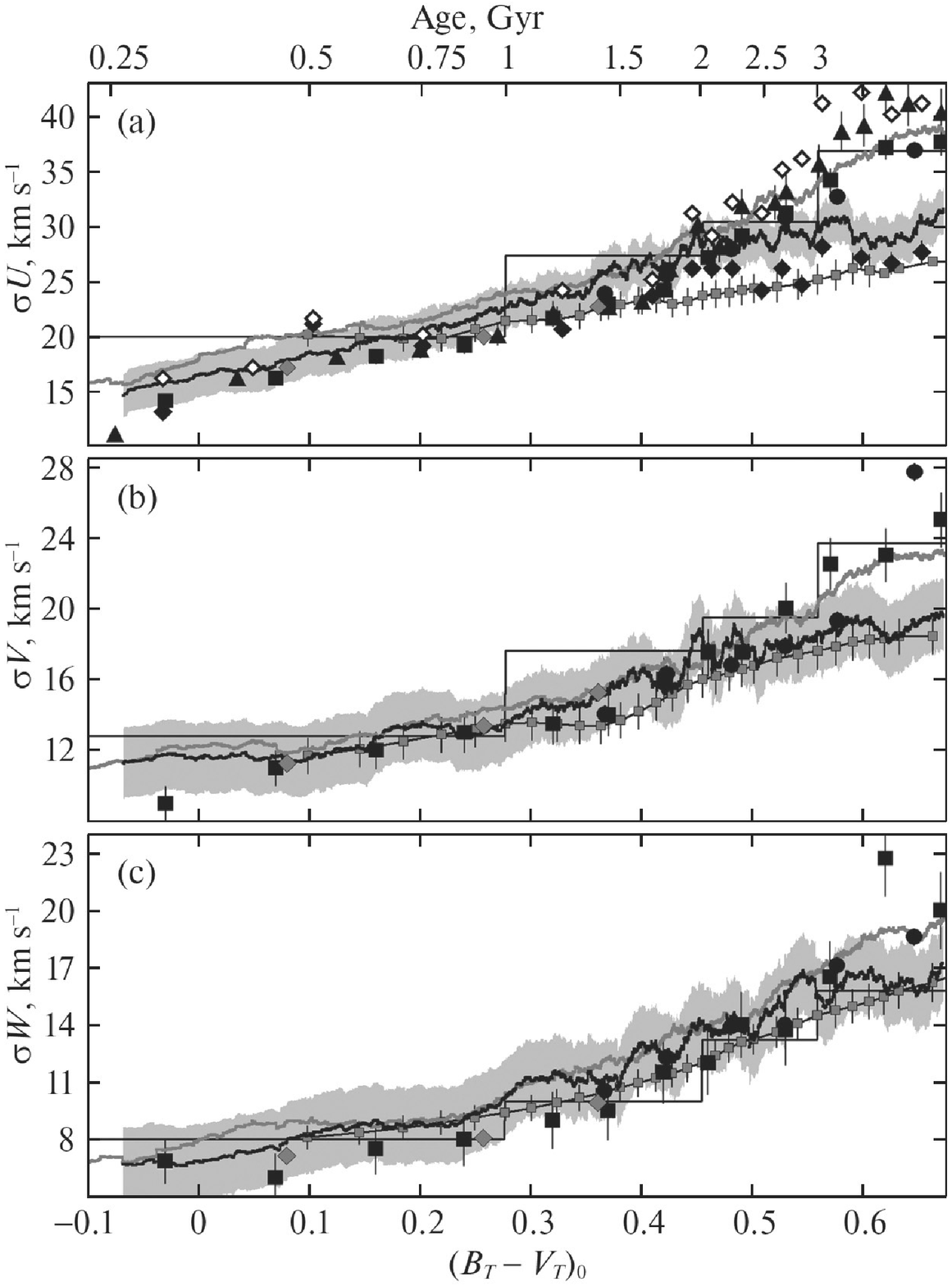}
\caption{Dispersions (a) $\sigma_U$, (b) $\sigma_V$, and (c) $\sigma_W$ versus $(B_T-V_T)_0$ and age for the studies 
specified in Table 1 before applying the reddening corrections. The black curve represents our result; the stepwise 
polygonal curve represents the BMG. The gray band $\pm2$ km s$^{-1}$ in width is shown for clarity.
}
\label{disp}
\end{figure}

\begin{figure}
\includegraphics{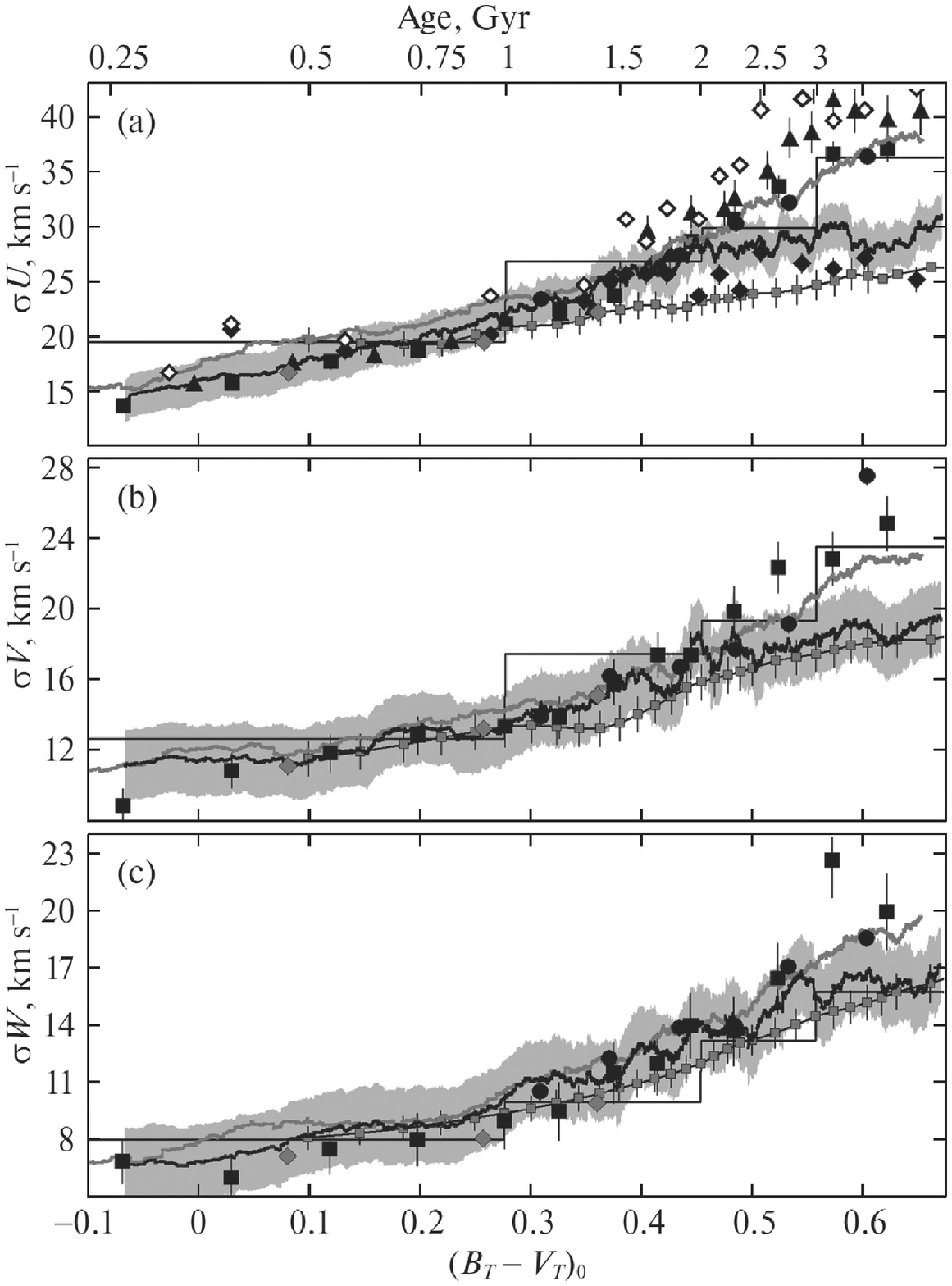}
\caption{Same as Fig. 1 after applying the reddening corrections.
}
\label{disp2}
\end{figure}

\begin{figure}
\includegraphics{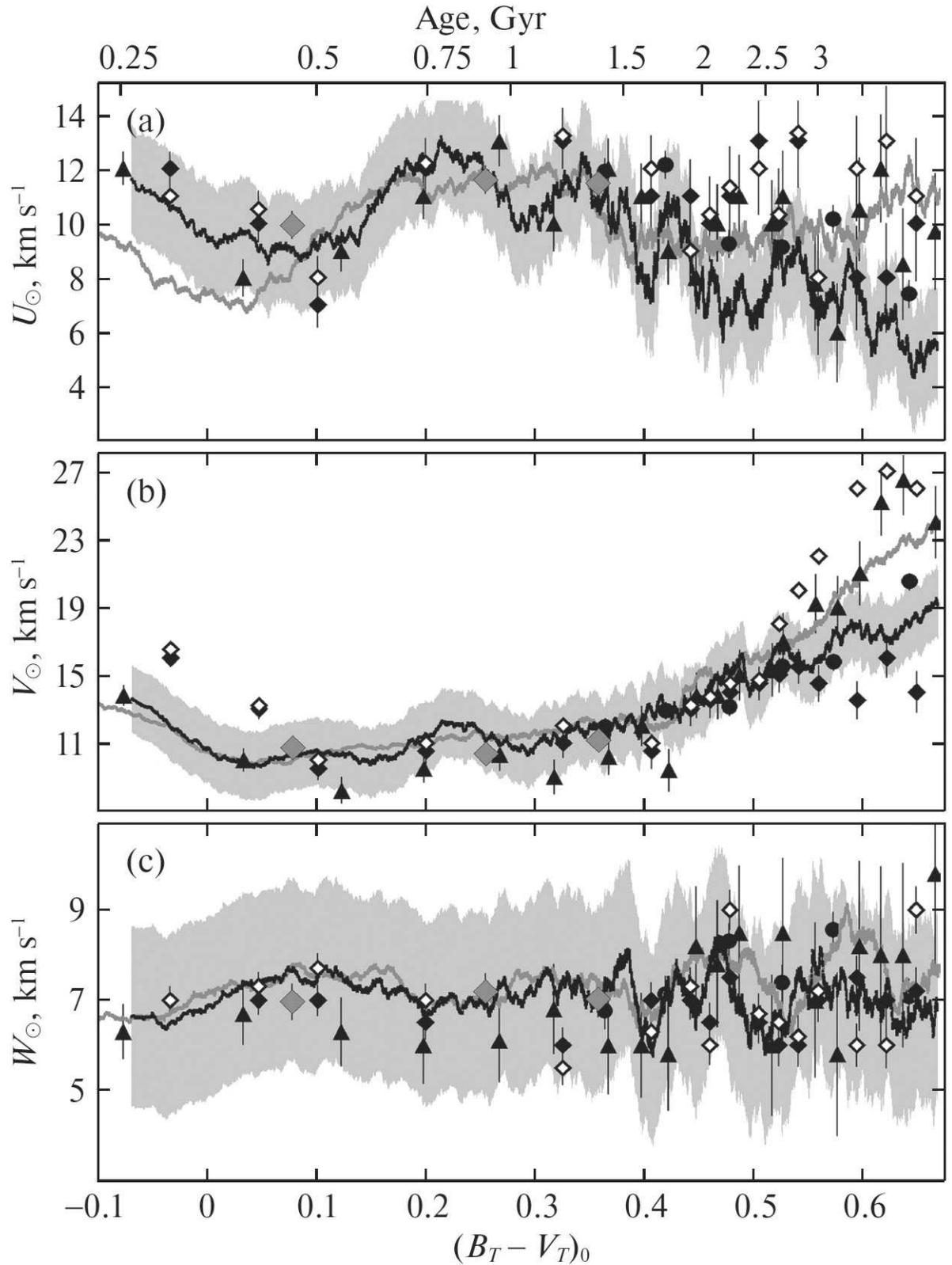}
\caption{Solar velocity components (a) $U_{\odot}$, (b) $V_{\odot}$, and (c) $W_{\odot}$ versus $(B_T-V_T)_0$ and age 
for the studies specified in Table 1 before applying the reddening corrections. The black curve represents our result. 
The gray band $\pm2$ km s$^{-1}$ in width is shown for clarity.
}
\label{uvw}
\end{figure}

\begin{figure}
\includegraphics{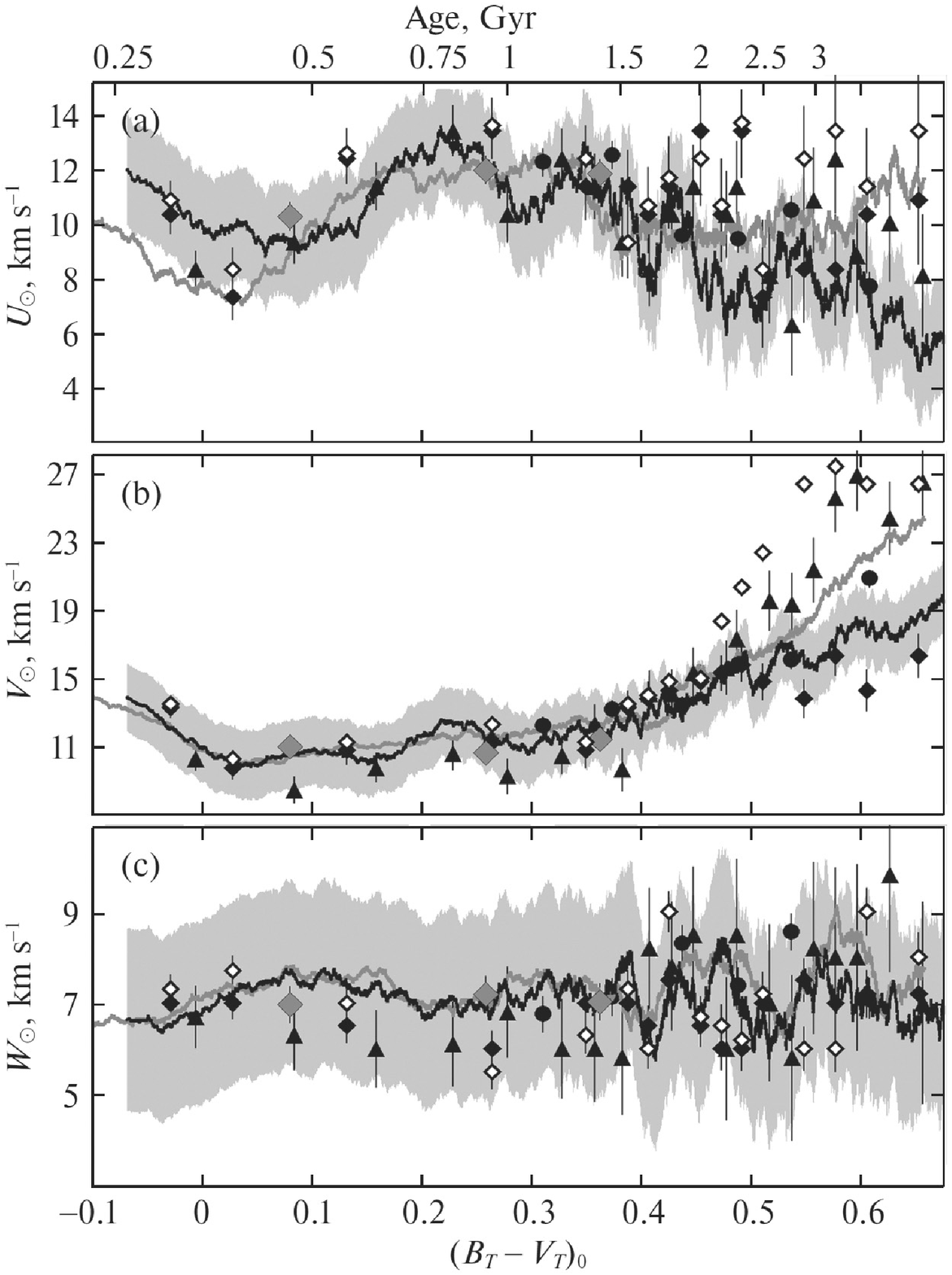}
\caption{Same as Fig. 3 after applying the reddening corrections.
}
\label{uvw2}
\end{figure}

\begin{figure}
\includegraphics{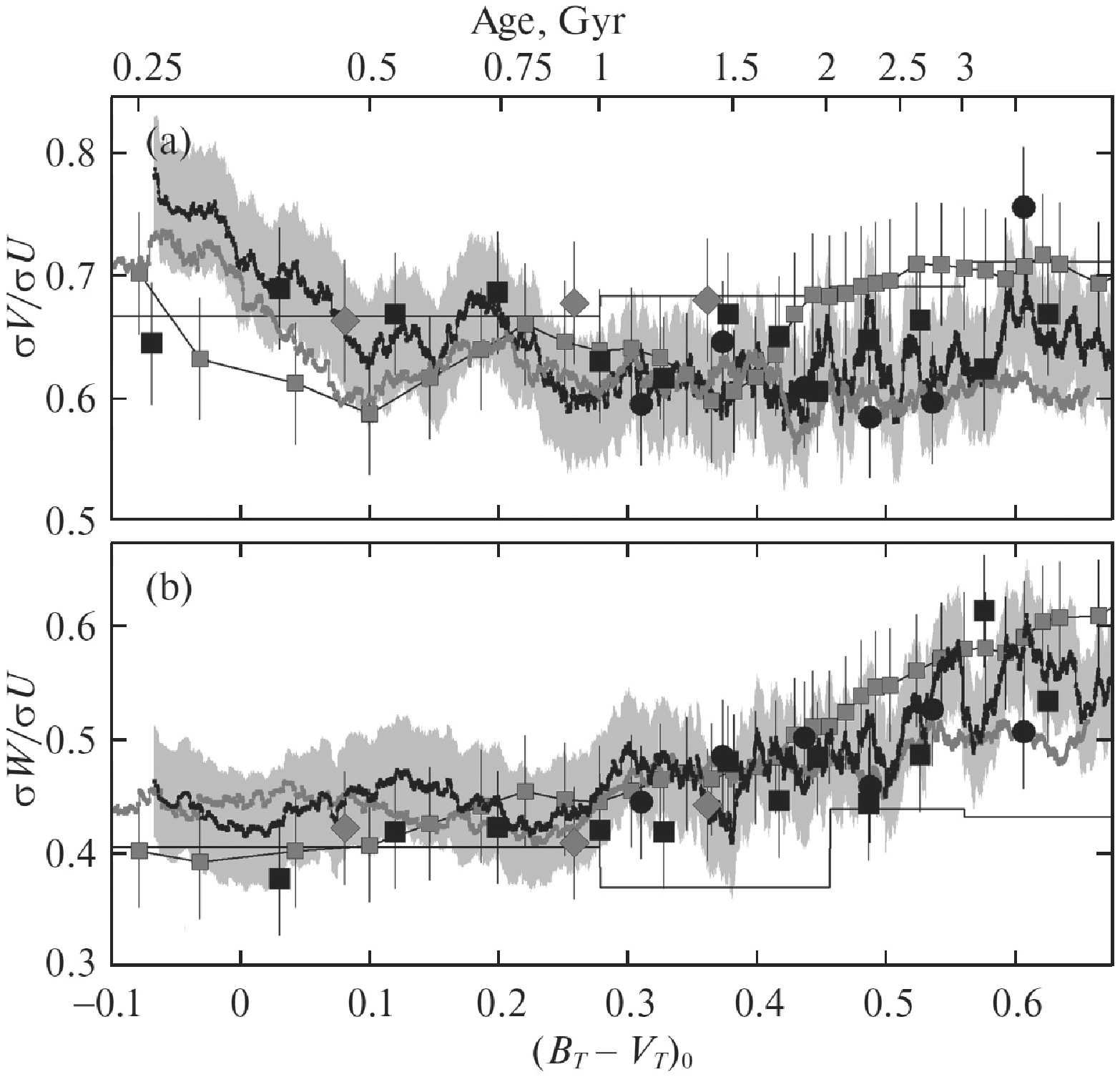}
\caption{Ratios (a) $\sigma_V/\sigma_U$ and (b) $\sigma_W/\sigma_U$ versus $(B_T-V_T)_0$ and age for the studies 
specified in Table 1 after applying the reddening corrections. The black curve represents our result; the stepwise 
polygonal curve represents the BMG. The gray band $\pm0.05$ in width is shown for clarity.
}
\label{susvsw}
\end{figure}

\section*{ORIGINAL DATA}

The distances $R$ for TGAS stars estimated by Astraatmadja and Bailer-Jones (2016) based on $\varpi$ and
their errors $\sigma(\varpi)$ from TGAS are used in this study. These $R$ slightly differ from the $R=1/\varpi$
estimate primarily because the Lutz-Kelker and Malmquist biases (Perryman 2009, p. 208) were taken into account.

The TGAS authors recommended to add an uncertainty of 0.3 mas, which describes the disregarded systematic errors
in $\varpi$ (Gaia 2016), to their formal parallax error $\sigma(\varpi)$. However, having
investigated their samples of clump giants from TGAS, Gontcharov (2017b) and Gontcharov and
Mosenkov (2017a) showed that the systematic errors of the TGAS parallaxes do not exceed 0.2 mas within
700 pc of the Sun. This value was added to $\sigma(\varpi)$ in this study.

Following the recommendation of Dehnen and Binney (1998), for this study we selected stars with
$\sigma(\varpi)/\varpi<0.1$. This corresponds to $R<330$ pc. 67\% and 95\% of the stars in the final sample are within
135 and 225 pc of the Sun, respectively. The median distance from the Sun for the sample stars changes
approximately linearly with $(B_T-V_T)_0$ from 170 pc at $(B_T-V_T)_0=-0.1^m$ to 60 pc at $(B_T-V_T)_0=0.7^m$. 
The median value is $\sigma(\varpi)/\varpi=0.045$. Since the sample stars are close to the Sun, this gives a
relative error in $R$ of 4.5\%.

The errors in $\mu_l\cos(b)$ and $\mu_b$ for all of the selected stars are less than 0.3 mas yr$^{-1}$. 
This is smaller than the analogous errors in Hipparcos approximately by a factor of 3. 
Such an increase in accuracy actually excludes the proper motions from the sources of errors
in $U$, $V$, $W$, and they are now determined by the errors in $R$ and $V_r$. Indeed, for the overwhelming
majority of stars 0.3 mas yr$^{-1}$ corresponds to a relative error much smaller than 4.5\%. Therefore,
the error in the tangential velocity components $V_l=4.74R\mu_l\cos(b)$ and $V_b=4.74R\mu_b$ is determined by the
error in $R$ and is less than 5\%. This corresponds to $\sigma(V_l)<1.25$, $\sigma(V_b)<1.25$ km s$^{-1}$, 
given that $V_l$ and $V_b$ do not exceed 25 km s$^{-1}$ in absolute value for the overwhelming majority of stars. 
The median error in $V_l$ and $V_b$ is 1 km s$^{-1}$.

The radial velocities for the TGAS stars under consideration are presented in the Radial Velocity
Experiment (RAVE) (Kunder et al. 2017) and LAMOST Spectroscopic Survey of Galactic Anti-Center
(Tian et al. 2015) projects, the mentioned CRVAD-2, PCRV, and GCS catalogues, and many small
catalogues. GCS was used in the PCRV after allowance made for the systematic errors, including
the error in $V_r$ as a function of $(B-V)$ reaching $\Delta V_r=1$ km s$^{-1}$ at $(B-V)=0.2^m$, which is especially
important for our study. The CRVAD-2 authors ignored these errors and simply averaged the PCRV
and GCS data for the corresponding stars. Thus, they actually transferred the GCS errors reduced by
half to CRVAD-2. The GCS, RAVE, and LAMOST catalogues refer mainly only to MS F--G stars. In
addition, RAVE and LAMOST cover only part of the sky and exhibit systematic errors in $V_r$ reaching
$\Delta V_r=1.5$ km s$^{-1}$ for RAVE (Gontcharov 2007) and $\Delta V_r=5.7$ km s$^{-1}$ for LAMOST (Tian et al. 2015).
Tian et al. (2015) showed that such errors could distort severely the kinematic characteristics. Thus,
the PCRV is the only catalogue of $V_r$ in which the systematic errors in $V_r$ were determined and taken
into account. Therefore, this study is restricted to using the PCRV. Other catalogues of $V_r$ are planned
to be used after investigating their systematic errors.

The PCRV contains the radial velocities for 35493 stars from the Hipparcos catalogue. The median
error in $V_r$ for the TGAS stars under consideration is 0.8 km s$^{-1}$, with $V_r$ being more accurate than
2 km s$^{-1}$ for the overwhelming majority of stars and more accurate than 5 km s$^{-1}$ for all stars.

Thus, the errors in $U$, $V$, and W for the stars under consideration are determined by the errors in $R$ and
$V_r$. From the formulas provided by Kulikovskii (1985, p. 74), given the errors in $V_l$, $V_b$, and $V_r$ of 
1.25, 1.25, and 2 km s$^{-1}$, respectively, for the overwhelming majority of stars under consideration (the median ones
are 1, 1, and 0.8 km s$^{-1}$), we find that the errors in $U$, $V$, and $W$ for them do not exceed 2.5 km s$^{-1}$
(the median error is 1.7 km s$^{-1}$). At present, this is a record high accuracy of the data used to calculate
the kinematic characteristics. Note that this progress was achieved mainly through the high accuracy of TGAS parallaxes.

However, the kinematic characteristics obtained in this study are the result of averaging 500 original
values. Therefore, the formal error of such a mean must be much smaller than the error in the velocity
of an individual star and smaller than the thickness of the black curve in the figures. Formally, the accuracy
of the results of this study is considerably higher than the typical accuracy of 0.4--2 km s$^{-1}$ in the
remaining studies under consideration, because $\varpi$ and $\mu$ from Hipparcos were replaced by their more
accurate values from TGAS. However, this study, along with all the remaining ones, may disregard
some systematic errors. The above accuracy was then overestimated. We note unresolved binary stars
(Gontcharov et al. 2001) as an obvious source of such systematic errors. Therefore, for clarity, the gray band
along the black curve in all figures indicates not the error of the mean for 500 stars but an upper limit for
the accuracy of the studies under consideration.

Most of the studies mentioned in Table 1 used $(B_T-V_T)$ or $(B_T-V_T)_0$ as an argument. Therefore,
$(B_T-V_T)_0$ was used as an argument for this study as well. 
The Tycho-2 photometry is accurate ($\sigma(B_T)<0.05^m$, $\sigma(V_T)<0.05^m$, the median accuracy is higher
than $0.02^m$) under the constraints $B_T<11^m$ and $V_T<10.7^m$. They were applied in this study. In addition,
the constraint $(B_T-V_T)_0<0.7^m$ was imposed on the sample to exclude the giants and MS stars of
late types. As a result, 9567 stars remained in the sample. Note that due to the small fraction of stars
with measured $V_r$, this sample, like any sample from Table 1, is not complete in the space under consideration.
Under the same selection conditions, but without restricting ourselves to the PCRV stars, we
would select 30509 and 97253 stars from Hipparcos and TGAS, respectively.

To exclude the stars that do not correspond to the thin-disk kinematics, for example, high-velocity
OB stars (Gontcharov and Bajkova 2013) and hot subdwarfs (Gontcharov et al. 2011), we applied the
criterion from Aumer and Binney (2009). A star was excluded if its total space velocity $(U^2+V^2+W^2)^{1/2}$
was larger than $3.5(\sigma_U^2+\sigma_V^2+\sigma_W^2)^{1/2}$ calculated, as
described previously, as a moving average for the corresponding window of 500 stars. Since the exclusion
of stars reduces the velocity ellipsoid, the process is repeated in iterations. We needed three instead of six
iterations in Aumer and Binney (2009). We rejected 24 stars instead of 55 in Aumer and Binney (2009);
9543 thin-disk stars remained in the final sample.

The reddening $E(B_T-V_T)=(B_T-V_T)-(B_T-V_T)_0$ and interstellar extinction $A_{V_T}$ were taken into
account in this study in accordance with the 3D reddening $E(J-Ks)$ map constructed by Gontcharov
(2017a) from 2MASS photometry (Skrutskie et al. 2006). Under the assumption of $E(B-V)=1.655E(J-Ks)$ 
this map gives a 3D $E(B-V)$ map and, together with the extinction-to-reddening $R_V=A_V/E(B-V)$ map from 
Gontcharov (2012a), a 3D $A_V$ map. Let us justify the use of precisely these reddening and extinction estimates.

\section*{REDDENING}

Since the visible range and extinction laws close to the law from Cardelli et al. (1989) with $A_V=3.1E(B-V)$ 
are considered in all of the above studies, the extinction estimate is uniquely related to the
reddening estimate. Below in this study we use everywhere the extinction law with $A_{B_T}=1.36A_V$, 
$A_{V_T}=1.06A_V$, $A_b=1.24A_V$ and $A_y=A_V$ adopted in the PARSEC database and very close to the extinction
law from Cardelli et al. (1989).

Let us compare the reddening estimates adopted in the studies from Table 1. G\'omez et al. (1997)
calculated the age from the positions of stars on the Hertzsprung--Russell (HR) diagram relative to
the isochrones by first estimating the effective temperature and metallicity from Str\"omgren photometry
or spectroscopic data and the reddening from Str\"omgren photometry. Apparently, their reddening
estimates may be deemed sufficiently accurate. Mignard (2000) did not need to take into account the
reddening, because the spectral classification of stars was used as an argument. Aumer and Binney (2009)
adopted zero reddening within 70 pc of the Sun and a subsequent growth in reddening $E(B-V)$ with
distance by $0.47^m$ kpc$^{-1}$. They noted that their revision of the reddening estimates compared to the preceding
studies changed significantly the estimates of some kinematic characteristics. Gontcharov (2012c)
estimated the reddening from the extinction ratio based on the 3D model from Gontcharov (2009b,
2012b) and the extinction-to-reddening ratio $R_V=A_V/E(B-V)$ based on the 3D map of spatial $R_V$
variations from Gontcharov (2012a). The new 3D reddening map from Gontcharov (2017a) shows that
Gontcharov (2012c) overestimated the reddening of blue stars. Aghajani and Lindegren (2013) adopted
the reddening from GCS calculated from Str\"omgren photometry. In the remaining studies the authors
did not specify how the reddening was taken into account. However, the star selection methods
specified by the authors allowed me to reconstruct (1) the composition of the samples for all of the
mentioned studies, except for G\'omez et al. (1997), (2) the distribution of the samples in $(B_T-V_T)_0$ and 
$(B_T-V_T)$, and (3) the mean reddening estimates adopted by the authors for each subsample. For
the subsamples of Mignard (2000) we calculated new means $\overline{(B_T-V_T)_0}$ slightly differing from the
original ones: $\overline{(B_T-V_T)_0}=0.08^m$, 0.267$^m$, and 0.375$^m$ for A0--A5, A5--F0, and F0--F5 stars, 
respectively.
The calibration $(B_T-V_T)_0=16(b-y)^3-7(b-y)^2+2.42(b-y)-0.08$ in accordance with the
PARSEC database was adopted for the results of Aghajani and Lindegren (2013). As a result, it was
established that Dehnen and Binney (1998) adopted reddening estimates close to those from Aumer and
Binney (2009), i.e., close to 0 for most stars, while Francis and Anderson (2009), apparently, did not
take into account the reddening at all.

\begin{figure}
\includegraphics{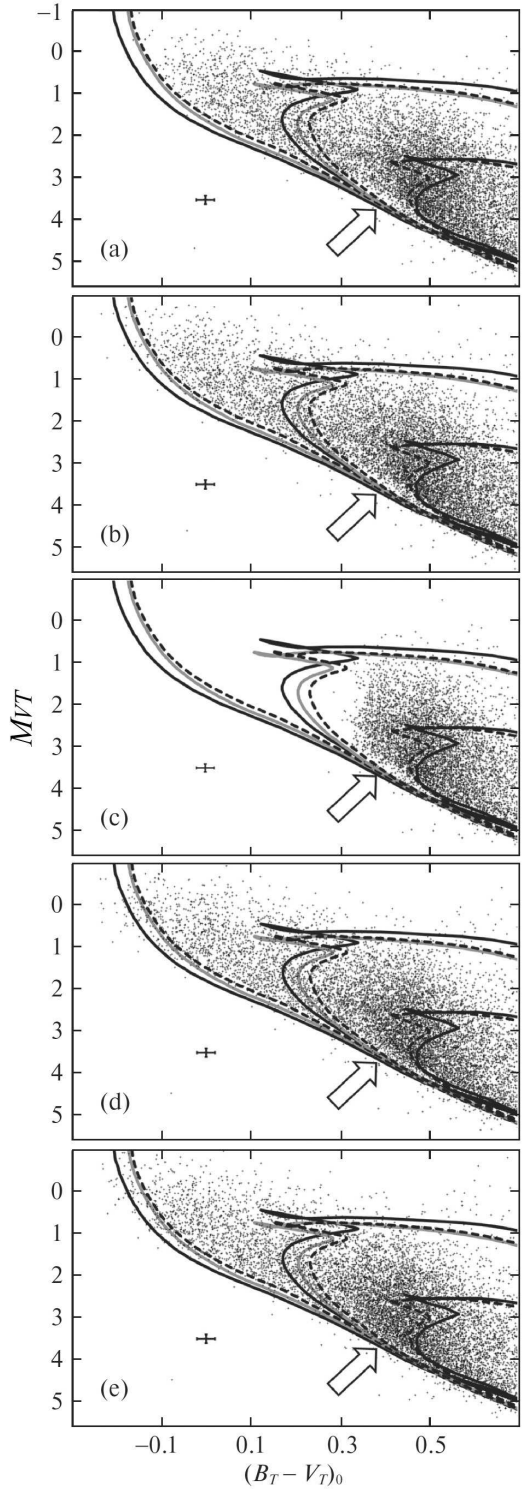}
\caption{The stars under consideration on the HR $(B_T-V_T)_0$ -- $M_{V_T}$ diagram uncorrected for reddening and 
extinction (a) and corrected in accordance with the estimates of Aumer and Binney (2009) (b), GCS (c), Arenou et al. 
(1992) (d), and Gontcharov (2012a, 2017a) (e). The lines indicate three sets of PARSEC (black solid curves) and MIST 
(with (black dashed curves) and without (gray solid curves) allowance for the stellar rotation) isochrones for 0.1, 1, 
and 3 Gyr (from left to right) and the metallicities specified in the text. The cross indicates the median errors of 
the data due to the photometric and parallax errors.
}
\label{hr}
\end{figure}

The positions of the 9543 stars under consideration on the HR $(B_T-V_T)_0$ -- $M_{V_T}$ diagram are shown
in Fig. 6, where $(B_T-V_T)_0$ was not corrected for reddening, while $M_{V_T}$ was not corrected for extinction
$A_{V_T}$ (Fig. 6a) or was corrected in accordance with the estimates of Aumer and Binney (2009)
(Fig. 6b), GCS (Fig. 6c) used by Aghajani and Lindegren (2013), Arenou et al. (1992) (Fig. 6d),
and Gontcharov (2012a, 2017a) (Fig. 6e). We chose precisely these reddening estimates, because
many other popular estimates cannot be used in the solar neighborhood $R<330$ pc under consideration,
as was pointed out by their authors. The estimates by Schlegel et al. (1998), Meisner
and Finkbeiner (2015), Green et al. (2015), and others are inapplicable here (Gontcharov 2017a; Gontcharov and
Mosenkov 2017b).

The lines in Fig. 6 indicate three sets of PARSEC and MIST isochrones for 0.1, 1, and 3 Gyr (from left
to right) and the following metallicities: $\mathbf Z=0.0152$, $\mathbf Y=0.2756$ 
(solar metallicity according to Bressan et al. (2012)) for PARSEC 0.1 and 1 Gyr and $\mathbf Z=0.0142$, $\mathbf Y=0.2738$
for 3 Gyr; MIST initial $\mathbf Y=0.27$, initial $\mathbf Z=0.0142$, actual $\mathbf Z=0.0156$. In contrast to
other models, MIST estimates the increase in metallicity as the star evolves and can take into account the
stellar rotation. When choosing the isochrones, we took into account the mean metallicity--age relation
for thin-disk stars (Haywood 2006). The isochrones for 0.1 Gyr may be deemed as the zero-age MS.

A sharp decrease in the number of stars at $(B_T-V_T)_0>0.5^m$ rightward of the 3--Gyr isochrones can
be seen in Fig. 6. This is how the magnitude limitation of the Hipparcos catalogue and, subsequently,
the PCRV manifested itself: these catalogues are much more complete with regard to B--A than F--G dwarfs 
(this can be seen, for example, in Fig. 6 from the PCRV description by Gontcharov (2006)).
Consequently, the kinematic characteristics for the range $0.5^m<(B_T-V_T)_0<0.7^m$ must be distorted
by selection more severely than those for the range $(B_T-V_T)_0<0.5^m$.

We see that for 0.1 Gyr the PARSEC and MIST isochrones almost coincide at $M_{V_T}>3^m$; otherwise
they diverge by no more than $\Delta (B_T-V_T)_0=0.05^m$.
Much of this divergence is caused by allowance for the stellar rotation in MIST.

YaPSI and BaSTI give no isochrones for $B_T$ and $V_T$ , but all four isochrones for 0.1 Gyr can be
compared on the effective temperature--luminosity diagram shown in Fig. 7. Here, $\mathbf Z=0.0162$, $\mathbf Y=0.28$
for YaPSI and $\mathbf Z=0.014$, $\mathbf Y=0.263$ for BaSTI.
The thick horizontal lines mark the luminosity range $-0.04<\log(L/L_{\odot})<2.27$ corresponding to the
range $0^m<M_{V_T}<5^m$ of interest to us. We see that here the YaPSI isochrone almost coincides with
PARSEC, while BaSTI is located approximately between PARSEC andMIST. The discrepancy between
the 0.1 Gyr isochrones is comparable to the errors of the photometry used and the error in the $E(B_T-V_T)$
estimates (for example, 0.03$^m$ for Arenou et al. (1992) and 0.04$^m$ for Gontcharov (2017a)). Therefore, the
positions of the stars relative to the isochrones can be used to choose the best source of reddening estimates.

\begin{figure}
\includegraphics{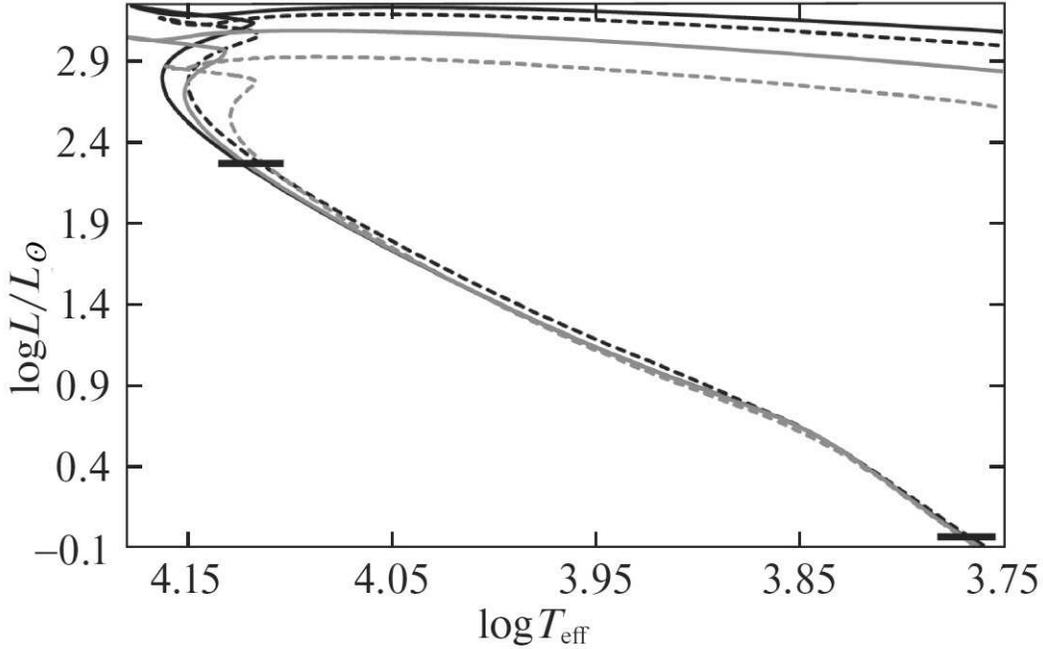}
\caption{PARSEC (black solid curve), MIST (black dashed curve), YaPSI (gray solid curve), and BaSTI
(gray dashed curve) isochrones on the HR effective temperature--luminosity diagram for an age of 0.1 Gyr
and the metallicities specified in the text. The thick horizontal lines mark the luminosity range of interest to us.
}
\label{tl}
\end{figure}

The arrow in Fig. 6 indicates the region of the HR diagram that is apparently most sensitive to the reddening
and extinction estimates: the isochrones here almost coincide, while the number of stars is great.
In this region on panel (a) the cloud of stars is offset from the isochrones, forming an unreasonable gap. It
is retained on panels (b) and (c), shrinks noticeably on panel (d), and disappears on panel (e). This means
that the mean reddening of the stars in this region of the diagram differs noticeably from zero, it was significantly
underestimated by Aumer and Binney (2009) and Aghajani and Lindegren (2013), was slightly underestimated
by Arenou et al. (1992), and was estimated most accurately by Gontcharov (2017a). The
same conclusion can be drawn from the upper left part of the graphs, where the variant with zero reddening,
and Aumer and Binney (2009) demonstrate a clearly insufficient number of youngest stars. The
mean reddening $E(B_T-V_T)$ for the stars under consideration is 0.063$^m$, 0.045$^m$, and 0.024$^m$, as inferred
by Gontcharov (2017a), Arenou et al. (1992), and Aumer and Binney (2009), respectively.

Adopting the reddening estimates from Gontcharov (2017a) as the most accurate ones, we will add the
corrections of the remaining studies for $E(B_T-V_T)$ from the column `$\Delta E(B_T-V_T)$' of Table 1 to $(B_T-V_T)_0$. 
The results of all studies corrected in this way are shown in Figs. 2, 4, and 5 (only the abscissas, but
not the ordinates change). When comparing Figs. 2 and 4 with Figs. 1 and 3, we see that the results come
significantly closer together.

\begin{figure}
\includegraphics{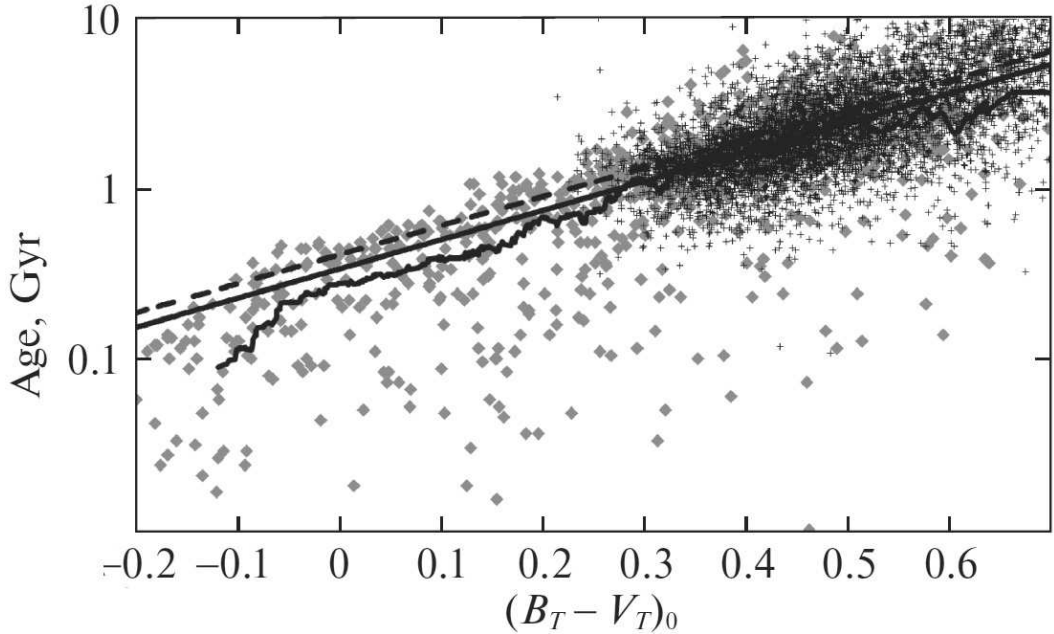}
\caption{Age in Gyr versus $(B_T-V_T)_0$ for the GCS stars (black dots, the fit by the black solid line) and 1000
TRILEGAL simulated stars with the properties of the stars of this study (gray diamonds, the black polygonal
curve is the moving average over 33 points). The black dashed curve is the fit of the previous GCS versions.
}
\label{age}
\end{figure}

\section*{AGE}

A correlation between the dereddened color and age for MS B-F stars was pointed out by the authors
of all studies from Table 1. The age estimates obtained by Casagrande et al. (2011) in the latest GCS version
from an analysis of the positions of GCS stars relative to the PARSEC and BaSTI isochrones on
the HR diagram are among the most accurate ones. Preliminary estimates of the metallicity of each star
were used for this purpose. Several age estimates were obtained for each star, depending on the set of
isochrones used and the method of calculation: the median, most probable, etc. However, they agree well
in the range $(B_T-V_T)_0<0.7^m$ we consider. Below we use the median age estimate from the BaSTI
isochrones. Figure 8 shows the relation between the age in Gyr and $(B_T-V_T)_0$ for the GCS stars. In this
case, $(B_T-V_T)_0$ was calculated from the reddening from the map of Gontcharov (2017a). The black solid
line indicates the following relation calculated for the GCS stars by the least-squares method:
\begin{equation}
\label{agecolor}
T=0.34e^{3.9(B_T-V_T)_0}
\end{equation}
The relation $T=0.41e^{3.9(B_T-V_T)_0}$ derived by Gontcharov (2012) from the previous GCS versions
(Nordstr\"om et al. 2004; Holmberg et al. 2007, 2009) is indicated in Fig. 8 by the dashed line. The change
of the coefficient in the relation is caused by the change of the calibrations in the latest GCS version.
Although GCS contains no blue stars, as can be seen from Fig. 8, extrapolating these relation to the
range $(B_T-V_T)_0<0.25^m$ gives quite plausible age estimates.

Figure 8 also shows the result of simulating the sample of TGAS stars considered in this study using
the TRIdimensional modeL of thE GALaxy (TRILEGAL; Girardi et al. 2005): 1000 simulated stars are
indicated by the gray diamonds. TRILEGAL also accurately reproduces the distribution of stars on the
HR diagram. The constraints $\sigma(\varpi)/\varpi<0.1$ and $V_T<10.7^m$ are seen to have deprived the sample of
many blue ($(B_T-V_T)_0<-0.1^m$) and red ($(B_T-V_T)_0>0.5^m$) stars.

For the TRILEGAL stars the dependence of the age on $(B_T-V_T)_0$ is indicated in Fig. 8 by the black
polygonal curve, the result of moving averaging over 33 points. There is good agreement between the
dependences for GCS and TRILEGAL, although the downward deflection of the polygonal curve at 
$(B_T-V_T)_0<-0.1^m$ and $(B_T-V_T)_0>0.5^m$ confirms that in these ranges the sample under consideration is
significantly incomplete with regard to the oldest stars and, hence, stars with a larger velocity dispersion.
Consequently, the sample under consideration is comparable in completeness and accuracy of kinematic
characteristics to other studies from Table 1 only in the range $-0.1^m<(B_T-V_T)_0<0.5^m$.

As a result, relation (1) was adopted everywhere in this study. In particular, it was used to calculate
$(B_T-V_T)_0$ from the age for the results of G\'omez et al. (1997). Relation (1) gives the age of an individual
star with a very low accuracy of 50--100\% and can be used only for the age statistics of a large number of stars.

The uniqueness of relation (1) allows us to plot the age scale on all figures in addition to the $(B_T-V_T)_0$
scale and to estimate the age dependences of the dispersions $\sigma_U$, $\sigma_V$ and $\sigma_W$ for the range of ages
from 0.23 to 2.4 Gyr, which corresponds to $-0.1^m<(B_T-V_T)_0<0.5^m$. The generally known proportionality
\begin{equation}
\label{sigmat}
\sigma\sim T^{\beta}.
\end{equation}
is confirmed in this study. An overview of previous $\beta$ determinations was given, for example, by Aumer
and Binney (2009) and Sharma et al. (2014): $0.27<\beta_{U,V}<0.39$ and $0.33<\beta_W<0.54$ are typical for the
thin disk.

The exponents $\beta$ found are compared with the results of other studies in Table 2. For the studies
from Table 1 two numbers separated by a slash are specified here: the original $\beta$ and $\beta$ for the range
$-0.1^m<(B_T-V_T)_0<0.5^m$ after applying the reddening corrections. The second value is more suitable
for a comparison with the results of this study. The exponents $\beta$ from the results of the first (Nordstr\"om
et al. 2004) and third (Holmberg et al. 2009) GCS versions are given only for comparison, because they
were actually replaced by the new values obtained by Aghajani and Lindegren (2013) from the same data.
Note that the previously specified systematic errors in $V_r$ from GCS did not manifest themselves in these results.

It can be seen from Table 2 that $\beta$ obtained in this study are close to the results of Gontcharov (2012c),
noticeably larger than $\beta$ from G\'omez et al. (1997), and slightly smaller than the remaining ones. Narrowing
the range of colors/ages and applying the reddening corrections reduced noticeably many of $\beta$,
eliminated the extreme values, and placed the results of Aumer and Binney (2009), Gontcharov (2012c),
Aghajani and Lindegren (2013), and this study in the narrow ranges $0.29<\beta_U<0.37$, $0.24<\beta_V<0.33$ 
and $0.32<\beta_W<0.40$ (the agreement of the results in Fig. 2 implies the agreement of $\beta$). Consequently,
an accurate allowance for reddening and the systematic errors of the original data is needed for the
conclusions about the kinematic characteristics to be reached.

\begin{table*}[!h]
\def\baselinestretch{1}\normalsize\normalsize
\caption[]{Exponents $\beta$ in the dependence (2).
}
\label{table2}
\[
\arraycolsep=0.05em
\begin{tabular}{lccc}
\hline
\noalign{\smallskip}
Reference & $\beta_U$ & $\beta_V$ & $\beta_W$ \\
\hline
\noalign{\smallskip}
G\'omez et al. (\cite[1997]{gomez}) & 0.15 / 0.15 & 0.24 / 0.22 & 0.35 / 0.32  \\
Nordstr\"om et al. (\cite[2004]{gcs}) & 0.31 & 0.34 & 0.47  \\
Holmberg et al. (\cite[2009]{gcs3}) & 0.39 & 0.40 & 0.53  \\
Aumer and Binney (\cite[2009]{aumer}) & 0.31 / 0.35 & 0.43 / 0.33 & 0.45 / 0.38  \\
Gontcharov (\cite[2012c]{apex}) & 0.31 / 0.29 & 0.24 / 0.24 & 0.35 / 0.32  \\  
Aghajani and Lindegren (\cite[2013]{aghajani}) & 0.38 / 0.37 & 0.51 / 0.32 & 0.47 / 0.40  \\
This study & 0.31 & 0.25 & 0.38 \\
\hline
\end{tabular}
\]
\end{table*}


\section*{OTHER SOURCES OF DISCREPANCIES}

In Table 2 for the results of G\'omez et al. (1997) $\beta$ are very small. The corresponding deviation of the results
of G\'omez et al. (1997) from the remaining ones can also be seen in Figs. 1 and 2: the gray squares are
shifted from the remaining results upward at $(B_T-V_T)_0<0.15^m$ and downward at $(B_T-V_T)_0>0.3^m$.
The results of Francis and Anderson (2009) for the thin-disk stars, i.e., the filled diamonds (in contrast
to the open ones) for $\sigma_U$ in Figs. 1a and 2a, look the same.

This is the result of an excessively stringent and color- or age-independent constraint in the selection
of thin-disk stars: G\'omez et al. (1997) rejected the stars with a space velocity $(U^2+V^2+W^2)^{1/2}>65$ km s$^{-1}$; 
Francis and Anderson (2009) rejected 5660 stars (28\% of their sample) with $(U^2+V^2+W^2)^{1/2}>84$ km s$^{-1}$. 
As a result, an admixture of stars that do not belong to the thin disk must be retained among the blue stars, 
while among the red stars, on the contrary, the fastest thin-disk stars were removed.

Our test showed that if the same constraints were imposed on the sample of Gontcharov (2012c), then
more than 10\% of the stars would be excluded from it. In this case, the dependences marked by the gray
curves in Figs. 1 and 2 are shifted downward and turn out to be close to the filled black diamonds (the results
of Francis and Anderson (2009)) and the gray squares (the results of G\'omez et al. (1997)). The results of
Francis and Anderson (2009), i.e., the positions of the filled diamonds relative to the open ones in Figs. 3b
and 4b, show that such an exclusion of the fastest stars also affects the dependence of $V_{\odot}$ on $(B_T-V_T)_0$, 
but only in the range $(B_T-V_T)_0>0.5^m$.

Thus, the deviation of the results of G\'omez et al. (1997) and Francis and Anderson (2009) for
the thin-disk stars from the results of other authors is explained by an incorrect selection criterion producing
strong selection. However, the thin-disk stars should be isolated. Otherwise, the admixtures
distort the results as much as, for example, the open diamonds in Figs. 1a, 2a, 3, and 4 deviate upward
from the remaining results. Apparently, in Figs. 2a and 4b at $(B_T-V_T)_0>0.4^m$ the results of Francis
and Anderson (2009) before and after the rejection of stars (open and filled diamonds) show the upper and
lower limits between which all plausible results must be located.

However, the deviation of the results of Francis and Anderson (2009) from the remaining ones at
$(B_T-V_T)_0=0.45^m$ and 0.49$^m$ in Fig. 4a requires a different explanation. The previously mentioned
systematic errors in $V_r$ from CRVAD-2 used by them can be the culprit here.

Another discrepancy can be seen in Figs. 1--4: at $-0.05^m<(B_T-V_T)_0<0.1^m$ and $0.5^m<(B_T-V_T)_0<0.7^m$ 
the results of this study deviate noticeably from the results of a similar study by
Gontcharov (2012c). Both were obtained using $V_r$ from the PCRV. This forces us to suggest either
selection, not due to the PCRV, or a systematic difference between the TGAS and Hipparcos parallaxes
and proper motions as a cause of the discrepancies. Indeed, in the range $0.5^m<(B_T-V_T)_0<0.7^m$ the
sample of Gontcharov (2012c) contains more than 1500 stars that are absent in TGAS, although they are
present in Hipparcos. The causes of their absence in TGAS are comparatively close dwarfs (mostly closer
than 70 pc) with very large proper motions, many of which are binary and variable stars. After their test
exclusion from the sample of Gontcharov (2012c), the dependences of the kinematic characteristics on
$(B_T-V_T)_0$ in the range $0.5^m<(B_T-V_T)_0<0.7^m$ coincided, within the error limits, with the results of
this study. Thus, it is confirmed once again that the results of this study are severely distorted by selection
in the range $0.5^m<(B_T-V_T)_0<0.7^m$.

However, this test exclusion of stars from the sample of Gontcharov (2012c) hardly affected the results
in the range $-0.05^m<(B_T-V_T)_0<0.1^m$, but, at the same time, the results of the two studies come
considerably closer together when using the same parallaxes and proper motions. Consequently, in this
range, i.e., for stars of late B subtypes and early A subtypes, TGAS gives parallaxes and proper motions
that differ systematically from those in Hipparcos. This is particularly clearly seen for the velocity component
$U_{\odot}$ in Fig. 4a. Thus, according to TGAS, these B- and A-type stars, which are observed at
Galactic longitudes of about 90$^{\circ}$ and 270$^{\circ}$ and, hence, belong to the Local spiral arm, move toward the
Galactic anticenter, on average, 3 km s$^{-1}$ faster and show a smaller scatter of velocities than that from the
Hipparcos data. The next, more complete Gaia data releases can answer the question of whether this is the
result of an increase in the Gaia accuracy compared to Hipparcos or, on the contrary, the result of zonal
systematic errors in TGAS.

The last systematic deviation from the main set of results is that in Figs. 3b and 4b the results of Dehnen
and Binney (1998) (black triangles) at $0.08^m<(B_T-V_T)_0<0.38^m$ show very small $V_{\odot}$. This may
stem from the fact that Dehnen and Binney (1998) did not use $V_r$.

Returning to the list of possible causes of the discrepancies between the results of the studies given
in the Introduction, we will note the following:
\begin{enumerate}
\item Dehnen and Binney (1998) used the data from the first Hipparcos version, while Aumer and Binney
(2009) used those from the new one. Otherwise these studies are very similar. In Fig. 2a we see a
coincidence between the results of these studies at $(B_T-V_T)_0<0.4^m$ and systematically smaller $\sigma_U$ in
Aumer and Binney (2009). Apparently, the transition from the first Hipparcos version to the new one manifested
itself in this way.
\item Aumer and Binney (2009) and Aghajani and Lindegren (2013) tested the influence of the radial velocities
on their results. In the range $0.25^m<(B_T-V_T)_0<0.5^m$ the dispersions derived with and without
$V_r$ almost coincided. However, in the range $0.5^m<(B_T-V_T)_0<0.7^m$ invoking $V_r$ in both studies systematically
increased the dispersions by 1--4 km s$^{-1}$, while $U_{\odot}$, $V_{\odot}$ and $W_{\odot}$ did not change.
\item The various effects of selection have been shown previously. We will only add that in combating
them, Dehnen and Binney (1998) proposed to impose strong magnitude constraints on the samples.
Francis and Anderson (2009) argued against the necessity of this by showing that the GCS, PCRV,
and CRVAD-2 catalogues of radial velocities produce selection that causes no significant biases of
the kinematic characteristics. At the same time, Gontcharov (2009a) showed that due to the strong
correlation of the apparent magnitude, the absolute magnitude, the absolute value of the proper motion,
the parallax, and the errors of these quantities for MS stars, any constraint or selection by any of these 
characteristics, in general, causes biases of the kinematic characteristics being determined. Consequently, they
are present in all of the results under consideration, including the results of Dehnen and Binney (1998).
Tests like those performed in this study are needed to reveal these biases.
\item For the results of Mignard (2000) and Aghajani and Lindegre (2013) $(B_T-V_T)_0$ were calculated
from the spectral type and the $(b-y)$ color, respectively. Nevertheless, it can be seen from Figs. 2, 4,
and 5 that their results agree well with the remaining ones. We will add that, in general, the uncertainties
of the calibrations (for example, $(B_T-V_T)_0$ from $(b-y)$) are currently smaller than the uncertainties of
the reddening estimates. Therefore, greater attention should be given to the latter.
\end{enumerate}

\section*{DISCUSSION}

The studies from Table 1 that showed no significant methodological flaws (Dehnen and Binney
1998; Mignard 2000; Aumer and Binney 2009; Gontcharov 2012; Aghajani and Lindegren 2013)
at $-0.1^m<(B_T-V_T)_0<0.5^m$ in Figs. 2 and 4 demonstrate agreement between themselves and
with the results of this study within $\pm2$ km s$^{-1}$. Their agreement within $\pm0.05$ can be seen in Fig. 5.
Both these quantities lie within the formal accuracy estimates for these studies. At the same time, the
variations of $\sigma_U$, $\sigma_V$, $\sigma_W$, $U_{\odot}$, $V_{\odot}$, $W_{\odot}$,  $\sigma_V/\sigma_U$ 
and $\sigma_W/\sigma_U$ in the same figures exceed noticeably $\pm2$ km s$^{-1}$ and $\pm0.05$. 
For example, a well-known growth of the dispersion with dereddened color and age can be seen in Fig. 2. 
Given the dependences (1) and (2), it can be described by the dependences derived by the least-squares method from 
all reliable data (in km s$^{-1}$):
\begin{equation}
\label{sigmabt}
\sigma_U=16.0e^{1.29(B_T-V_T)_0}, \sigma_V=10.9e^{1.11(B_T-V_T)_0}, \sigma_W=6.8e^{1.46(B_T-V_T)_0}.
\end{equation}

They correspond to
\begin{equation}
\label{beta}
\beta_U=0.33, \beta_V=0.285, \beta_W=0.37.
\end{equation}
The results of all these best studies fit into the range $\pm0.05$ relative to these values. However, these values
of $\beta$ refer only to comparatively young stars in the ranges of colors $-0.1^m<(B_T-V_T)_0<0.5^m$ and
ages from 0.23 to 2.4 Gyr. Therefore, we can draw only cautious conclusions, especially pending more
accurate Gaia data. However, it can be argued that with a high probability
\begin{equation}
\label{beta3}
\beta_V<\beta_U<\beta_W, 
\end{equation}
with $\beta_U$ and $\beta_W$ being close to 1/3 and $\beta_V$ being between 1/3 and 1/4. All three exponents are quite
far from the popular estimates of 0.4--0.5 in previous studies. A modern overview of $\beta$ and the corresponding
causes of the increase in velocity dispersion with age was given by Aumer et al. (2016). The influence of giant 
molecular clouds, the spiral pattern, the bar, satellite galaxies, a nonuniform distribution
of dark matter, and, in general, any source of gravity causing a deviation of the Galactic gravitational
field in the solar neighborhood from the axisymmetric one are among the causes. Yet another cause is
a decrease in the initial velocity dispersion of stars during their birth as the Galaxy evolves due to the
decrease in the total gas mass in the disk. Aumer et al. (2016) simulated the influence of many of these
causes on $\beta$ and concluded that the spiral pattern, the bar, and giant molecular clouds play a key role.
When comparing the results of Aumer et al. (2016) with $\beta$ from this study, the discrepancy in $\beta_W$ is
particularly pronounced. Only one of the models of Aumer et al. (2016), Y1s2, gives $\beta_W=0.41$ comparable
to that obtained in this study. The remaining models give much larger values. Consequently,
the results of this study argue for the parameters of the Galaxy that provide a low value of $\beta_W$. Aumer
et al. (2016) pointed out these parameters: a more constant star formation rate causing a slow decrease
in the mass of giant molecular clouds as the Galaxy evolves; giant molecular clouds play an important,
but not exceptional role in the growth of the velocity dispersion, including $\sigma_W$; a comparatively large mass
(including the dark halo) lies far from the Galactic plane. They also proposed a scenario providing a
minimum $\beta_W$ when the condition (5) is met precisely for comparatively young stars, as in this study: the
spiral pattern, at a secondary role of giant molecular clouds, causes an increase in $\sigma_U$ and $\sigma_V$ only 
until some moment, and giant molecular clouds then continue to increase $\sigma_W$, turning the orbits of stars
with a large eccentricity and a small inclination to the Galactic plane into orbits with a small eccentricity
and a large inclination. Aumer et al. (2016) pointed out that the ratio $\sigma_W/\sigma_U$ is a measure of the influence
of giant molecular clouds or the spiral pattern that increase the stellar velocity dispersions, respectively,
in all directions or only along the Galactic plane. The gradual growth in $\sigma_W/\sigma_U$ with stellar age found in
this study corresponds to the assumption made by Aumer et al. (2016) about an increase in the relative
importance of the spiral pattern compared to the importance of clouds, which are expended to a
considerable extent on star formation.

$U_{\odot}$, $V_{\odot}$ and $W_{\odot}$ variations with an amplitude of 7, 9, and 2.7 km s$^{-1}$, respectively, 
can be seen in Figs. 4a--4c. The latter value is small enough to adopt the mean and the range of variations around it in
km s$^{-1}$ for the entire range $-0.1^m<(B_T-V_T)_0<0.5^m$:
\begin{equation}
\label{wodot}
\overline{W_{\odot}}=7.15\pm1.35.
\end{equation}
In the same range the $U_{\odot}$ and $V_{\odot}$ variations in km s$^{-1}$ can be fitted by the dependences derived 
from all reliable studies:
\begin{equation}
\label{uodot}
U_{\odot}=782(B_T-V_T)_0^4-698(B_T-V_T)_0^3+225.6(B_T-V_T)_0^2-23.5(B_T-V_T)_0+10.38
\end{equation}
\begin{equation}
\label{vodot}
V_{\odot}=1228(B_T-V_T)_0^4-1214(B_T-V_T)_0^3+319(B_T-V_T)_0^2-8.8(B_T-V_T)_0+8.34.
\end{equation}
These relations do not reflect the high-frequency variations that, for example, are visible in the results of
Dehnen and Binney (1998) and this study: the local maximum of $U_{\odot}$ and $V_{\odot}$ at $(B_T-V_T)_0\approx0.23^m$, 
another local maximum of $U_{\odot}$ at $(B_T-V_T)_0\approx0.35^m$, and two local minima of $U_{\odot}$ at 
$(B_T-V_T)_0\approx0.29^m$ and $0.41^m$. These studies are completely independent,
did not use any common data, and Dehnen and Binney (1998) did not use $V_r$ at all. Therefore, these
extrema can be real and need a further study.

The dependence for $V_{\odot}$ is a manifestation of the well-known asymmetric drift, i.e., the dependence
of the rotation velocity of stars around the Galactic center on their age (Perryman 2009, p. 305). It
can be seen from Fig. 4 that at $(B_T-V_T)_0>0.04^m$, i.e., at an age older than 0.4 Gyr, $V_{\odot}$ increases with
$(B_T-V_T)_0$, though not monotonically. The theoretical linear dependence of $V_{\odot}$ on the age (Francis and
Anderson 2009) is not confirmed: in addition to it, there are high-frequency variations. It is also obvious
that stars of approximately the same, accurately determined age should be used for any estimates of
the peculiar solar motion relative to the group of stars (for a discussion, see Gontcharov 2012c).

In Fig. 5 the $\sigma_V/\sigma_U$ and $\sigma_W/\sigma_U$ variations reach $\pm0.1$ and $\pm0.07$, respectively. 
$\sigma_V/\sigma_U$ and $\sigma_W/\sigma_U$ change their behavior in unison: at $(B_T-V_T)_0<0.25^m$, i.e., at an age 
younger than 0.9 Gyr, $\sigma_V/\sigma_U$ decreases with $(B_T-V_T)_0$ and $\sigma_W/\sigma_U=0.42\pm0.05$ is constant, 
while at $0.25^m<(B_T-V_T)_0<0.5^m$ $\sigma_V/\sigma_U=0.63\pm0.05$ is constant and $\sigma_W/\sigma_U$
increases with $(B_T-V_T)_0$ to $\sigma_W/\sigma_U\approx0.55$. For comparison, Famaey et al. (2005) found stabilization
for branch giants with an age older than 3 Gyr at $\sigma_V/\sigma_U=0.65$ and $\sigma_W/\sigma_U=0.50$, i.e., close to
our values. However, their simulations gave 0.79 and 0.55, where the former differs greatly from the empirical values.

It can be seen from Fig. 2 that, on the whole, $\sigma_U$ and $\sigma_V$ adopted in the BMG (stepwise curve) at
$(B_T-V_T)_0>0^m$ describe satisfactorily the growth of these dispersions with dereddened color and age,
while for $\sigma_W$ they are clearly underestimated. At $(B_T-V_T)_0<0^m$ the BMG values were clearly overestimated.
In the BMG the thin-disk stars were separated by age into seven populations with discrete values of their kinematic 
characteristics. This gives the stepwise pattern of the dependences for the BMG in Figs. 1, 2, and 5. A comparison of
these dependences with the remaining results shows that the kinematic characteristics of the stars under
consideration change gradually rather than abruptly, as in the BMG.

\section*{CONCLUSIONS}

In recent years, the original data traditionally used to study the kinematics of stars near the Sun
have been updated substantially. The Gaia DR1 TGAS catalogue gave the most accurate parallaxes
and proper motions, the PARSEC, MIST, YaPSI, and BaSTI databases gave improved theoretical
isochrones, a number of studies with reddening and interstellar extinction estimates near the Sun have
appeared. These data were tested in our study when calculating the dependences of kinematic characteristics
on the dereddened color and the age correlating with it. The age was taken from the Geneva-
Copenhagen survey in agreement with the estimates of the theoretical TRILEGAL database. The radial
velocities were taken from the PCRV, which covers the entire sky and is most free from the systematic
errors. The parallax and radial velocity errors were shown to introduce the main uncertainty into the results.

Based on the criterion $(B_T-V_T)_0<0.7^m$, for our study we chose a region of the HR diagram that is
kinematically interesting and sensitive to the data accuracy: B--F types of the MS. We selected 9543
thin-disk stars with accurate Tycho-2 photometry and parallaxes $\sigma(\varpi)/\varpi<0.1$. This provided a median
relative accuracy of 1.7 km s$^{-1}$ for the velocity components $U$, $V$, and $W$, the highest one among all
studies of kinematic characteristics.

We considered the variations of $\sigma_U$, $\sigma_V$, $\sigma_W$, $U_{\odot}$, $V_{\odot}$, $W_{\odot}$, 
$\sigma_V/\sigma_U$ and $\sigma_W/\sigma_U$ with $(B_T-V_T)_0$ in comparison with the key results of other authors 
obtained after the appearance of the Hipparcos catalogue. Various tests showed a significant effect of
selection on the kinematic characteristics, especially outside the range $-0.1^m<(B_T-V_T)_0<0.5^m$. For
example, the excessively stringent and color- and age-independent criteria for the selection of thin-disk
stars in the studies of G\'omez et al. (1997) and Francis and Anderson (2009) explain the deviation
of their results from the remaining ones. The kinematic characteristics calculated from the TGAS data
outside the range $-0.1^m<(B_T-V_T)_0<0.5^m$ are also significantly biased due to selection in TGAS.
However, in the range $-0.1^m<(B_T-V_T)_0<0.5^m$ the results obtained only from the proper motions or
radial velocities are close.

The positions of the stars under consideration on the HR diagram with PARSEC, MIST, YaPSI, and
BaSTI isochrones when using different reddening and extinction estimates showed the following: (1) the
isochrones are in agreement with one another and (2) the reddening and extinction estimates are erroneous
in some kinematic studies. The estimates based on the 3D reddening map from Gontcharov (2017a)
in combination with the 3D extinction-to-reddening map from Gontcharov (2012a) were recognized to
be the most reliable ones. The colors of the stars used in other studies were corrected for an inaccurate
reddening correction.

As a result, the dependences of the kinematic characteristics on $(B_T-V_T)_0$ obtained by various
authors approached significantly one another and the results of our study. The scatter of $\sigma_U$, 
$\sigma_V$, $\sigma_W$, $U_{\odot}$, $V_{\odot}$ and $W_{\odot}$ in reliable studies in the range
$-0.1^m<(B_T-V_T)_0<0.5^m$ fits into a $\pm2$ km s$^{-1}$ corridor, while the scatter of $\sigma_V/\sigma_U$ 
and $\sigma_W/\sigma_U$ fits into $\pm0.05$. This corresponds to the upper bound of
the declared accuracy of the studies under consideration. This allowed the variations of the kinematic
quantities as a function of $(B_T-V_T)_0$ and age in relations (3)--(8) to be reliably determined. These
variations make the determination of the solar motion relative to the stars without specifying their age meaningless.

\section*{ACKNOWLEDGMENTS}

I am grateful to the referees for their useful remarks. In this study I used resources from the Strasbourg
Astronomical Data Center (http://cdsweb.ustrasbg. fr) and Hipparcos/Tycho results. In this
study I used data from the Gaia mission of the European Space Agency (https://www.cosmos.esa.int/gaia) 
processed by the Gaia Data Processing and Analysis Consortium (DPAC, https://www.cosmos.esa.int/web/gaia/dpac/consortium).



\begin{thebibliography}{99}

\bibitem{planck} A. Abergel, P. A. R. Ade, N. Aghanim, M. I. R. Alves, G. Aniano, C. Armitage-Caplan, M. Arnaud, et al.
(Planck Collab.), Astron. Astrophys. \textbf{571}, A11 (2014).

\bibitem{arenou} F.~Arenou, M.~Grenon, and A.~G\'omez, Astron. Astrophys. \textbf{258}, 104 (1992).

\bibitem{bailer3} T.-L.~Astraatmadja and C.A.L.~Bailer-Jones, Astrophys. J. \textbf{833}, 119 (2016).

\bibitem{aumer} M.~Aumer and J.J.~Binney, Mon. Not. R. Astron. Soc. \textbf{397}, 1286 (2009).

\bibitem{aumer2016} M.Aumer, J.~Binney, and R.~Sch\"onrich, Mon. Not. R. Astron. Soc. \textbf{462}, 1697 (2016).

\bibitem{bovy} J.~Bovy, Mon. Not. R. Astron. Soc. \textbf{468}, 63 (2017).

\bibitem{bressan} A.~Bressan, P.~Marigo, L.~Girardi, B.~Salasnich, C.~Dal Cero, S.~Rubele, and A.~Nanni,
Mon. Not. R. Astron. Soc. \textbf{427}, 127 (2012), http://stev.oapd.inaf.it/cmd.

\bibitem{gaia} A.G.A.~Brown, A.~Vallenari, T.~Prusti, J.H.J.~de~Bruijne, F.~Mignard, R.~Drimmel, C.~Babusiaux,
C. A.L.~Bailer-Jones, et al. (Gaia Collab.), Astron. Astrophys. \textbf{595}, A2 (2016).

\bibitem{cardelli} J.A.~Cardelli, G.C.~Clayton, and J.S.~Mathis, Astrophys. J. \textbf{345}, 245 (1989).

\bibitem{casagrande} L.~Casagrande, R.~Sch\"onrich, M.~Asplund, S.~Cassisi, I.~Ram\'irez, J.~Mel\'endez, T.~Bensby, 
and S.~Feltzing, Astron. Astrophys. \textbf{530}, A138 (2011).

\bibitem{choi} J.~Choi, A.~Dotter, C.~Conroy, M.~Cantiello, B.~Paxton, B.D.~Johnson, Astrophys. J. \textbf{823}, 102 (2016).

\bibitem{bmg2} M.A.~Czekaj, A.C.~Robin, F.~Figueras, X.~Luri, and M.~Haywood, Astron. Astrophys. \textbf{564}, A102 (2014).

\bibitem{db} W.~Dehnen and J.~Binney, Mon. Not. R. Astron. Soc. \textbf{294}, 429 (1998).

\bibitem{mist} A.~Dotter, Astrophys. J. Supp. Ser. \textbf{222}, 8 (2016), http://waps.cfa.harvard.edu/MIST/

\bibitem{hip} ESA, \emph{Hipparcos and Tycho catalogues} (ESA, 1997).

\bibitem{famaey} B.~Famaey, A.~Jorissen, X.~Luri, M.~Mayor, S.~Udry, H.~Dejonghe, and C.~Turon, Astron. Astrophys. \textbf{430}, 165 (2005).

\bibitem{francis} C.~Francis and E.~Anderson, New Astronomy \textbf{14}, 615 (2009).

\bibitem{trilegal} L.~Girardi, M.A.T.~Groenewegen, E.~Hatziminaoglou, L.~da~Costa, Astron. Astrophys. \textbf{436}, 895, (2005), http://stev.oapd.inaf.it/cgi-bin/trilegal

\bibitem{gomez} A.E.~G\'omez, S.~Grenier, S.~Udry, M.~Haywood, L.~Meillon, V.~Sabas, A.~Sellier, and D.~Morin,
\emph{Proceedings of the ESA Symp. ``Hipparcos - Venice 97'', ESA SP-402, Ed.~B.~Battrick,
ESA Publications Division, c/o ESTEC, Noordwijk, The Netherlands}, 621 (1997).

\bibitem{pcrv} G.A.~Gontcharov, Astron. Lett. \textbf{32}, 759 (2006).

\bibitem{pcrvrave} G.A.~Gontcharov, Astron. Lett. \textbf{33}, 390 (2007).

\bibitem{model} G.A.~Gontcharov, Astron. Lett. \textbf{35}, 638 (2009a).

\bibitem{gould} G.A.~Gontcharov, Astron. Lett. \textbf{35}, 780 (2009b).

\bibitem{rgb} G.A.~Gontcharov, Astron. Lett. \textbf{37}, 707 (2011).

\bibitem{rv} G.A.~Gontcharov, Astron. Lett. \textbf{38}, 12 (2012a).

\bibitem{av} G.A.~Gontcharov, Astron. Lett. \textbf{38}, 87 (2012b).

\bibitem{apex} G.A.~Gontcharov, Astron. Lett. \textbf{38}, 771 (2012c).

\bibitem{g17} G.A.~Gontcharov, Astron. Lett. \textbf{43}, 472 (2017a).

\bibitem{g2017} G.A.~Gontcharov, Astron. Lett. \textbf{43}, 545 (2017b).

\bibitem{orbit} G.A.~Gontcharov and A.T.~Bajkova, Astron. Lett. \textbf{39}, 689 (2013).

\bibitem{gm2017} G.A.~Gontcharov and A.V.~Mosenkov, Mon. Not. R. Astron. Soc. \textbf{470}, L97 (2017a).

\bibitem{gm2017big} G.A.~Gontcharov and A.V.~Mosenkov, Mon. Not. R. Astron. Soc. \textbf{472}, 3805 (2017b).

\bibitem{pmfs} G.A.~Gontcharov, A.A.~Andronova, O.A.~Titov, and E.V.~Kornilov, Astron. Astrophys. \textbf{365}, 222 (2001).

\bibitem{sdwd} G.A.~Gontcharov, A.T.~Bajkova, P.N.~Fedorov, and V.S.~Akhmetov, Mon. Not. R. Astron. Soc. \textbf{413}, 1581 (2011).

\bibitem{green} G.M.~Green, E.F.~Schlafly, D.P.~Finkbeiner, H.-W.~Rix, N.~Martin, W.~Burgett, P.W.~Draper, et al.,
Astrophys. J. \textbf{810}, 25 (2015), http://argonaut.skymaps.info

\bibitem{haywood} M.~Haywood, Mon. Not. R. Astron. Soc. \textbf{371}, 1760 (2006).

\bibitem{tycho2} E.~H\o g, C.~Fabricius, V.V.~Makarov, S.~Urban, T.~Corbin, G.~Wycoff, U.~Bastian, P.~Schwekendiek,
and A.~Wicenecet, Astron. Astrophys. \textbf{355}, L27 (2000).

\bibitem{gcs2} J.~Holmberg, B.~Nordstr\"om, and J.~Andersen, Astron. Astrophys. \textbf{475}, 519 (2007).

\bibitem{gcs3} J.~Holmberg, B.~Nordstr\"om, and J.~Andersen, Astron. Astrophys. \textbf{501}, 941 (2009).

\bibitem{crvad} N.V.~Kharchenko, R.-D.~Scholz, A.E.~Piskunov, S.~R\"oser, and E.~Schilbach, Astronomische Nachrichten \textbf{328}, 889 (2007).

\bibitem{kulik} P. G. Kulikovskii, Stellar Astronomy (Nauka, Moscow, 1985) [in Russian].

\bibitem{kunder} A.~Kunder, G.~Kordopatis, M.~Steinmetz, T.~Zwitter, P.J.~McMillan, L.~Casagrande, H.~Enke, J.~Wojno,
et al., Astron. J. \textbf{153}, 75 (2017), http://www.rave-survey.org

\bibitem{hip2} F.~van Leeuwen, Astron. Astrophys. \textbf{474}, 653 (2007).

\bibitem{2015ApJ...798...88M} A.M.~Meisner and D.P.~Finkbeiner, Astrophys. J. \textbf{798}, 88 (2015).

\bibitem{michalik} D.~Michalik, L.~Lindegren, and D.~Hobbs, Astron. Astrophys. \textbf{574}, A115 (2015).

\bibitem{mignard} F.~Mignard, Astron. Astrophys. \textbf{354}, 522 (2000).

\bibitem{gcs} B.~Nordstr\"om, M.~Mayor, J.~Andersen, J.~Holmberg, F.~Pont, B.R.~J\o rgensen, E.H.~Olsen, S.~Udry, et al.,
Astron. Astrophys. \textbf{418}, 989 (2004).

\bibitem{parenago} P. P. Parenago, A Course in Stellar Astronomy (GITTL,Moscow, 1954) [in Russian].

\bibitem{paxton2011} B.~Paxton, L.~Bildsten, A.~Dotter, F.~Herwig, P.~Lesaffre, and F.~Timmes, Astrophys. J. Supp. Ser. \textbf{192}, 3 (2011).

\bibitem{per} M.~Perryman, \emph{Astronomical Applications of Astrometry} (Cambridge Univ. Press, Cambridge, 2009).

\bibitem{basti} A.~Pietrinferni, S.~Cassisi, M.~Salaris, and F.~Castelli, Astrophys. J. \textbf{612}, 168 (2004), http://basti.oa-teramo.inaf.it

\bibitem{robin} A.C.~Robin, C.~Reyl{\'e}, S.~Derri{\`e}re, and S.~Picaud, Astron. Astrophys. \textbf{409}, 523 (2003).

\bibitem{sfd} D.J.~Schlegel, D.P.~Finkbeiner, and M.~Davis, Astrophys. J. \textbf{500}, 525 (1998).

\bibitem{sharma} S.~Sharma, J.~Bland-Hawthorn, J.~Binney, K.C.~Freeman, M.~Steinmetz, C.~Boeche, O.~Bienaym\'e, B.K.~Gibson,
G.F.~Gilmore, et al., Astrophys. J. \textbf{793}, 51 (2014).

\bibitem{2mass} M.F.~Skrutskie, R.M.~Cutri, R.~Stiening, M.D.~Weinberg, S.~Schneider, J.M.~Carpenter, C.~Beichman, R.~Capps, et al.,
Astron. J. \textbf{131}, 1163 (2006), http://www.ipac.caltech.edu/2mass/releases/allsky/index.html

\bibitem{yapsi} F.~Spada, P.~Demarque, Y.-C.~Kim, T.S.~Boyajian, J.M.~Brewer, Astrophys. J. \textbf{838}, 161 (2017), http://www.astro.yale.edu/yapsi/
	
\bibitem{tian} H.-J.~Tian, C.~Liu, J.L.~Carlin, Y.-H.~Zhao, X.-L.~Chen, Y.~Wu, G.-W.~Li, Y.-H.~Hou, and Y.~Zhang, Astrophys. J. \textbf{809}, 145 (2015).



\end{thebibliography}
\end{document}